\documentclass[
   ,final            % use final for the camera ready runs
  ,numberedheadings % uncomment this option for numbered sections
  ]
  {aipproc}

\usepackage{rotating}

\layoutstyle{6x9}

\newcommand{\etal}{{\em et al.}}
\newcommand{\nim}{{Nucl. Instrum. Meth.}}
\newcommand{\dd}{\ensuremath{\mathrm{d}}}
\newcommand{\pizero}{{\ensuremath{\pi^{0}}}}
\newcommand{\gflash}{{\tt GFLASH}}
\newcommand{\geant}{{\tt GEANT}}
\newcommand{\geantthree}{{\tt GEANT3}}
\newcommand{\EoP}{{\ensuremath{\langle E/p\rangle}}}

\begin{document}

\title{Performance of the CDF Calorimeter Simulation in Tevatron Run~II}

\classification{07.05.Tp, 07.20.Fw, 13.85.-t}
\keywords{CDF, GFLASH, calorimeter simulation}
\author{Pedro~A.~Movilla~Fern\'{a}ndez  \\
 ({\small \em for the CDF Collaboration})\vspace*{-2pt}}
 { address={Lawrence Berkeley National Laboratory,
   One Cyclotron Road, Berkeley, CA 94720, U.S.A.
 }
}

%%%%%%%%%%%%%%%%%%%%%%%%%%%%%%%%%%%%%%%%%%%%
%% Abstract
%%%%%%%%%%%%%%%%%%%%%%%%%%%%%%%%%%%%%%%%%%%%

\begin{abstract}
The CDF experiment is successfully collecting data from
$\mathrm{p}\bar{\mathrm{p}}$ collisions at the Tevatron in Run~II. As
the data samples are getting larger, systematic uncertainties due to
the measurement of the jet energy scale assessed using the calorimeter
simulation have become increasingly important. In many years of
operation, the collaboration has gained experience with \gflash, a
fast parametrization of electromagnetic and hadronic showers used for
the calorimeter simulation.  We present the performance of the
calorimeter simulation and report on recent improvements based on a
refined {\em in situ} tuning technique. The central calorimeter
response is reproduced with a precision of 1-2\%.
\end{abstract}

\maketitle

%%%%%%%%%%%%%%%%%%%%%%%%%%%%%%%%%%%%%%%%%%%%
%% Introduction
%%%%%%%%%%%%%%%%%%%%%%%%%%%%%%%%%%%%%%%%%%%%

\section{Introduction}

Since the start of Run~II in 2001, the Collider Detector at Fermilab
(CDF)~\cite{CDF-tdr} has collected over 1\,fb$^{-1}$ of data of
$\mathrm{p}\bar{\mathrm{p}}$ collisions at 1.96~TeV center-of-mass
energy. In a variety of resulting publications, the simulation of the
calorimeter has proved to be a crucial element for the precision
measurement of physical observables, like the mass of the top quark,
since it appears as one of the keys to control the jet energy scale
systematics. The CDF calorimeter simulation is based on
\gflash~\cite{grindhammer:1990}, a FORTRAN package used for the fast
simulation of electromagnetic and hadronic showers. It is embedded in
a \geantthree\ framework~\cite{brun:1993} as part of the whole
detector simulation and has various advantages w.r.t. the detailed
\geant\ shower simulation: In CDF it is up to 100 times faster, and it
can be flexibly tuned.  The calorimeter simulation was initially tuned
to test beam data and has been improved due to a steadily refined {\em
in situ} tuning using samples of isolated charged particles, which
recently became available over a remarkably extended momentum range up
to 40\,GeV.  Here we report on the \gflash\ performance effective for
current CDF physics publications and on ongoing improvements
contributing to the ambitious Run~II physics program.

\vspace*{-2pt}

%%%%%%%%%%%%%%%%%%%%%%%%%%%%%%%%%%%%%%%%%%%%
%% CDF Calorimetry
%%%%%%%%%%%%%%%%%%%%%%%%%%%%%%%%%%%%%%%%%%%%

\section{CDF Calorimetry} 

\begin{figure}
  \includegraphics[height=.3\textheight]{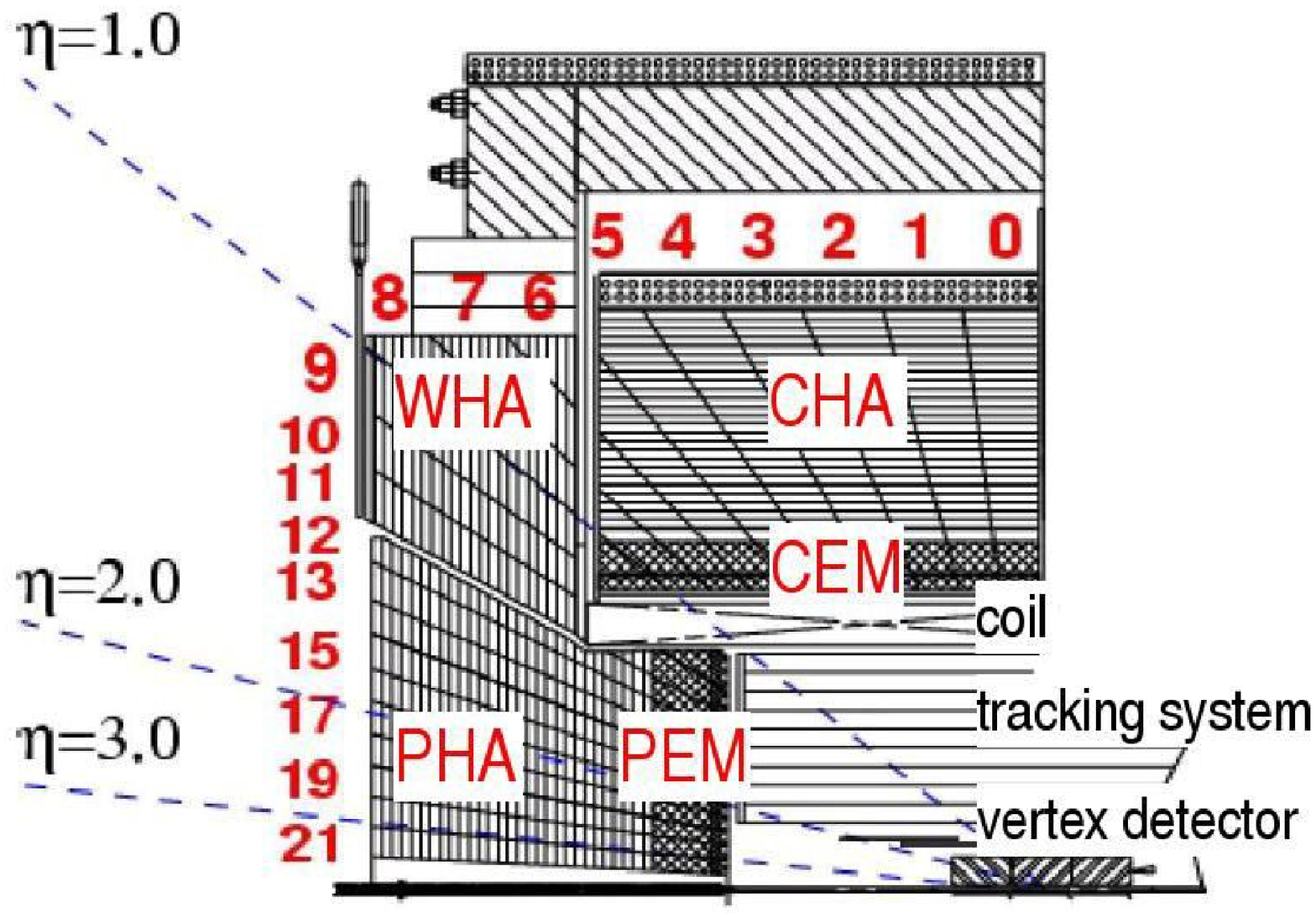}
  \includegraphics[height=.28\textheight]{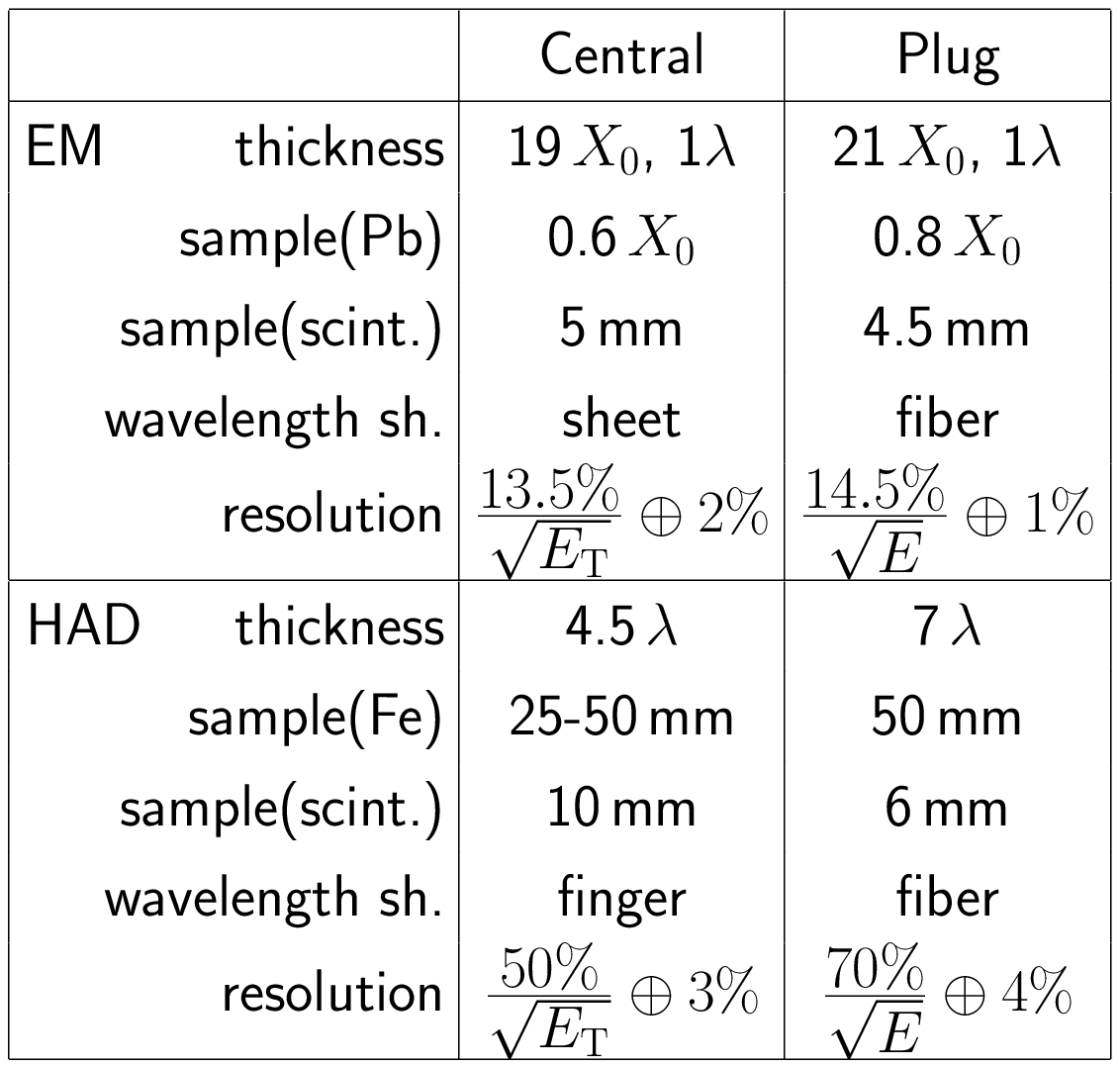} \caption{
  \label{fig:cdfcal}
   {\em Left:} Quadrant view of the CDF calorimeter showing the
   electromagnetic ({\sf CEM}, {\sf PEM}) and hadronic compartments
   ({\sf CHA}, {\sf WHA}, {\sf PHA}).
   {\em Right:} Sampling structure and energy resolutions for the
   central ({\sf CEM}, {\sf CHA}) and the plug part ({\sf PEM}, {\sf
   PHA}). \vspace*{-14pt}}
\end{figure}

CDF is a general-purpose charged and neutral particle detector with a
calorimeter and a tracking system. The calorimeter has a central and a
forward section subdivided into a total of five compartments
(Fig.\,\ref{fig:cdfcal}): the central electromagnetic,
CEM~\cite{CDF-cem}, the central hadronic, CHA~\cite{CDF-had}, the plug
electromagnetic, PEM~\cite{CDF-plug}, the plug hadronic,
PHA~\cite{CDF-plug}, and the wall hadronic, WHA~\cite{CDF-had}.  The
calorimeter is of sampling type, with lead (iron) absorbers for the
electromagnetic (hadronic) compartments, scintillating tiles and
wavelength shifters. It is subdivided into 44 projective tower groups,
each group made of 24 wedges (partially 48 in the plug part)
circularly arranged around the Tevatron beam and pointing to the
nominal event vertex.  The central and plug part together cover the
pseudorapidity range $|\eta|<3.6$. The calorimeter encloses a tracking
system consisting of a vertex detector and a cylindrical drift
chamber, both situated within a 1.4\,T solenoid magnet.  It provides a
precise measurement of single charged particle momenta serving as an
energy reference for the calorimeter simulation tuning.

% Fig. CDF Detector
\vspace*{-5pt}

%%%%%%%%%%%%%%%%%%%%%%%%%%%%%%%%%%%%%%%%%%%%
%% GFLASH
%%%%%%%%%%%%%%%%%%%%%%%%%%%%%%%%%%%%%%%%%%%%

\section{GFLASH in a Nutshell} 

The CDF simulation uses \geant\ to propagate particles from the main
interaction point through the detector volume. A shower in \gflash\ is
initiated when a particle undergoes the first inelastic interaction in
the calorimeter. \gflash\ treats the calorimeter as one effective
medium using \geant\ geometry and material information. \gflash\ is
ideal for calorimeter modules which may have a complicated but
repetitive sampling structure.

Given an incident particle energy, $E_\mathrm{inc}$, the visible
energy in the active medium,
% Eq: visible energy
\begin{equation} \label{eq:evis}  \vspace*{-4pt}
\dd  E_\mathrm{vis} (\mathbf{r}) 
   = E_\mathrm{inc} \hat{m} \left[
     \frac{\hat{e}}{\hat{m}} c_\mathrm{em} f_\mathrm{em}(\mathbf{r}) 
   + \frac{\hat{h}}{\hat{m}} c_\mathrm{had} f_\mathrm{had}(\mathbf{r})
     \right]
   \dd \mathbf{r} \;, \vspace*{-1pt}
\end{equation}
is calculated according to the sampling fractions for electrons
($\hat{e}$) and hadrons ($\hat{h}$) relative to the sampling fraction
for minimum ionizing particles ($\hat{m}$), taking their relative
fractions $c_\mathrm{em}$ and $c_\mathrm{had}$ of energy deposited in
the active medium into account. $f_\mathrm{em}(\mathbf{r})$ and
$f_\mathrm{had}(\mathbf{r})$ are electromagnetic and hadronic spatial
energy distributions of the form $f(\mathbf{r})=\frac{1}{2\pi}\,L(z)
\cdot T(r,\,z)$, factorizing into a longitudinal profile $L(z)$, which
is a function of the shower depth $z$, and a lateral profile
$T(r,\,z)$, which depends on $z$ and on the radial distance $r$ from
the shower center. The showers are treated as azimuthally symmetric.

The longitudinal electromagnetic profiles are assumed to follow a
Gamma distribution,
% Eq: longitudinal electron/photons
\begin{equation} \label{eq:long-em}
 L_\mathrm{em}(x) = \frac{x^{\alpha-1} e^{-x}}{\Gamma(\alpha)}\;,
\end{equation}
where $x=\beta z$ and $z$ measured in units of radiation lengths
$X_0$. $\alpha$ and $\beta$ are correlated parameters generated using
two Gaussians, whose means and widths are free parameters subject to
tuning, and a correlation matrix hardwired in \gflash.  Longitudinal
hadronic shower profiles are a superposition of three shower classes:
% Eq: longitudinal hadrons
\begin{eqnarray} 
 L_\mathrm{had}(x) &=& f_\mathrm{dep} 
 \left[ c_\mathrm{h}\mathcal{L}_\mathrm{h}(x_\mathrm{h}) 
      + c_\mathrm{f}\mathcal{L}_\mathrm{f}(x_\mathrm{f}) 
      + c_\mathrm{l}\mathcal{L}_\mathrm{l}(x_\mathrm{l}) 
 \right]  \label{eq:long-had-a}\;,\\
\mathcal{L}_i(x_i)&=&\frac{x_i^{\alpha_i-1} e^{-x_i}}{\Gamma(\alpha_i)}
 \;\;,\;x_i=\beta_i z_i 
 \;\;\;\;\;\;\;\;\;\;(i=\mathrm{h}\,,\mathrm{f}\,,\mathrm{l}) \label{eq:long-had-b}
\;.
\end{eqnarray}
$\mathcal{L}_\mathrm{h}$ is a purely hadronic component
($z_\mathrm{h}$ given in units of absorption lengths
$\lambda_0$). $\mathcal{L}_\mathrm{f}$ accounts for the component
induced by neutral pions from a first inelastic interaction
($[z_\mathrm{f}]=X_0$), and $\mathcal{L}_\mathrm{l}$ originates from
neutral pions occurring in later stages of the shower development
($[z_\mathrm{l}]=\lambda_0$). Each subprofile in
Eq.\,(\ref{eq:long-had-b}) is characterized by an individual
correlated pair ($\alpha$, $\beta$) analogously to
Eq.\,(\ref{eq:long-em}).  The coefficients $c_i$ are the relative
probabilities of the three classes expressed in terms of the fraction
of showers containing a neutral pion ($f_\pizero$) and the fraction of
showers with a neutral pion in later interactions
($f^\mathrm{l}_\pizero$):
% Eq: longitudinal hadrons 2
\begin{equation} \label{eq:long-had-c}
c_\mathrm{h}=1-f_\pizero\;,\;\;
c_\mathrm{f}=f_\pizero(1-f^\mathrm{l}_{\pizero})\;,\;\;
c_\mathrm{l}=f_\pizero f^\mathrm{l}_{\pizero}\;.
\end{equation}
The global factor $f_\mathrm{dep}$ is the fraction of deposited energy
w.r.t. the energy of the incident particle.  When a shower is
generated, correlations between all parameters are properly taken into
account.  In total, the longitudinal shower profile is described by 18
independent parameters for the hadronic part (the means and widths of
the $\alpha$'s, $\beta$'s, and of the fractions $f_\mathrm{dep}$,
$f_\pizero$ and $f^\mathrm{l}_{\pizero}$), and four parameters for the
purely electromagnetic part.

The lateral energy profile at a given shower depth $z$ has the
functional form
% Eq: lateral 
\begin{equation} \label{eq:lat}
 T(r) = \frac{2 r R_0^2}{(r^2 + R_0 ^2)^2 }\;.
\end{equation}
The free quantity $R_0$ is given in units of Moli\`ere radius (for
electromagnetic) or absorption lengths (for hadronic showers),
respectively.  $R_0$ is an approximate log-normal distribution with a
mean and a variance parametrized as a function of the incident
particle energy $E_\mathrm{inc}$ and shower depth $z$. The mean is
given by
% Eq: lateral 2
\begin{eqnarray} \label{eq:lat-b}
  \langle R_0(E_\mathrm{inc},z)\rangle    
   & = & \left[ R_1 + (R_2 - R_3 \ln E_\mathrm{inc}) z \right]^n \;, 
\end{eqnarray}
where $n$=1(2) for the hadronic (electromagnetic) case. The spread of
hadronic showers increases linearly with shower depth $z$ and
decreases logarithmically with $E_\mathrm{inc}$. Both shower types
have their own set of adjustable parameter values (the $R_i$ plus
three independent parameters for the variance), thus giving a total of
12 parameters.

After generating the profiles, \gflash\ distributes the incident
particle energy in discrete interval steps following the longitudinal
profile, and then deposits {\em energy spots} in the simulated
calorimeter volume according to the lateral profile. The number of
energy spots is smeared to account for sampling fluctuations. The
visible energy is obtained by integrating over the energy spots and
applying the relative sampling fractions Eq.\,(\ref{eq:evis}).

%%%%%%%%%%%%%%%%%%%%%%%%%%%%%%%%%%%%%%%%%%%%
%% Gflash Tuning Method
%%%%%%%%%%%%%%%%%%%%%%%%%%%%%%%%%%%%%%%%%%%%

\section{Gflash Tuning Method}

The tuning of \gflash\ mostly relies on test beams of electrons and
charged pions with energies between 8 and 230 GeV~\cite{currat:2002},
and has been refined during Run~II using the calorimeter response of
single isolated tracks in the momentum range 0.5-40\,GeV/c measured
{\it in situ}. The tuning of hadronic showers follows a four-step
procedure.

\paragraph{Adjusting the MIP peak} Reproducing the response of minimum 
ionizing particles (MIP) is the first step since it serves as a
reference for all other responses, see Eq.\,(\ref{eq:evis}). The
position and width of the simulated MIP peak is tuned to 57\,GeV test
beam pions in the electromagnetic compartments and involves the
adjustment of charge collection efficiencies, which is handled by
\geant\ at this stage of the simulation.

\paragraph{Setting the hadronic energy scale and shape} Next, the shape 
of the individual energy distributions in the electromagnetic and
hadronic compartment, the sum of both, and the hadronic response of
MIP-like particles in the hadronic compartment are adjusted. The
tuning is based on 57\,GeV test beam pions and involves the most
sensitive longitudinal parameters, the means and widths of
$\alpha_\mathrm{l}$ and $\beta_\mathrm{l}$ (steering the shower
component induced by \pizero's from later interactions), of
$\beta_\mathrm{h}$ (which is related to the purely hadronic component), see
Eq.\,(\ref{eq:long-had-b}), and of the fractions $f_\mathrm{dep}$ and
$f_\mathrm{\pizero}$ in Eqn.\,(\ref{eq:long-had-a}) and
(\ref{eq:long-had-c}).

\paragraph{Fixing the energy dependence} After the longitudinal 
profile has been adjusted at one energy point, the other test beam
samples are employed to parametrize the energy dependence of the
parameters, which is typically logarithmic. At this point, also the
inclusion of {\em in situ} tracks is important in order to provide a
robust extrapolation into the important low energy region where test
beam data are not available. A functional form
$f(E_\mathrm{inc})=a+b\tanh(c\ln E_\mathrm{inc}+d)$ is used to
describe the dependence of the fractions in
Eqn.\,(\ref{eq:long-had-a}) and (\ref{eq:long-had-c}) on the incident
particle energy $E_\mathrm{inc}$.  Distinct parametrizations in the
central and plug calorimeter were introduced to account for their
different sampling structure. Initial Run~II tunes involved samples of
single isolated tracks up to 5\,GeV/c in the central and plug
part. Recently, the energy evolution of $f_\mathrm{dep}$ and the
relative sampling fractions, $\hat{e}/\hat{m}$ and $\hat{h}/\hat{m}$,
have been tuned using the total and MIP-like responses of {\em in
situ} tracks up to 40\,GeV/c (Sec.\,\ref{sec:insitu}).

\paragraph{Tuning the lateral profile} The lateral profile is treated 
almost independently from longitudinal profile details and is tuned
solely with Run~II data. The parameters $R_i$ of
Eq.\,(\ref{eq:lat-b}), which have been initially adjusted to describe
the profiles in the energy range 0.5-5.0\,GeV/c (using default values
provided by the H1 collaboration at higher energies), are now tuned up
to particle momenta of 40\,GeV/c (Sec.\,\ref{sec:insitu}).

%%%%%%%%%%%%%%%%%%%%%%%%%%%%%%%%%%%%%%%%%%%%
% Impact on CDF Jet Energy Scale
%%%%%%%%%%%%%%%%%%%%%%%%%%%%%%%%%%%%%%%%%%%%

\section{Impact on CDF Jet Energy Scale}

% Fig: E/p NIM + JES
\begin{figure}
  \includegraphics[height=.25\textheight, width=.49\textwidth] 
  {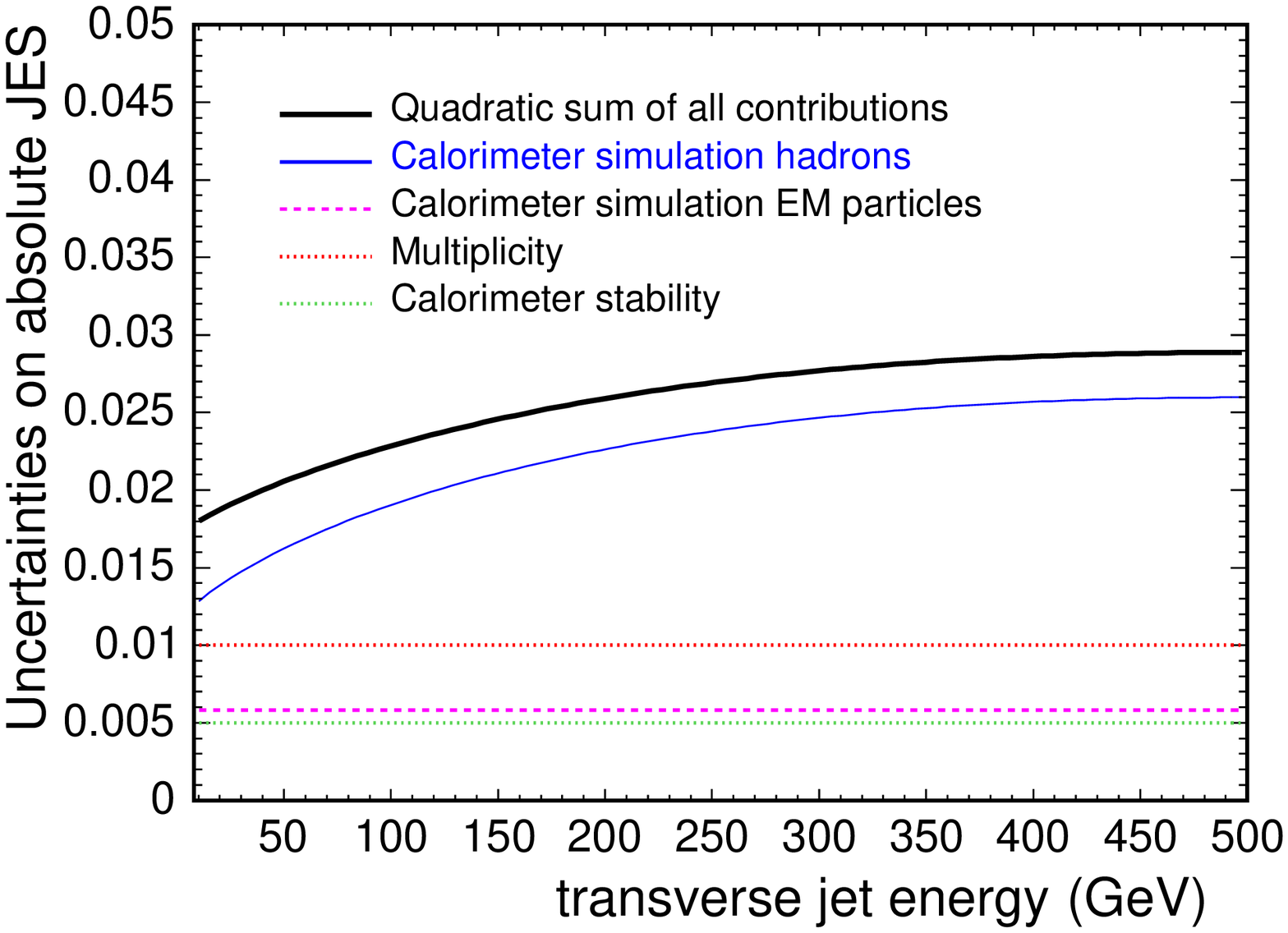}
  \hspace*{5pt}
  \includegraphics[height=.25\textheight, width=.49\textwidth]
  {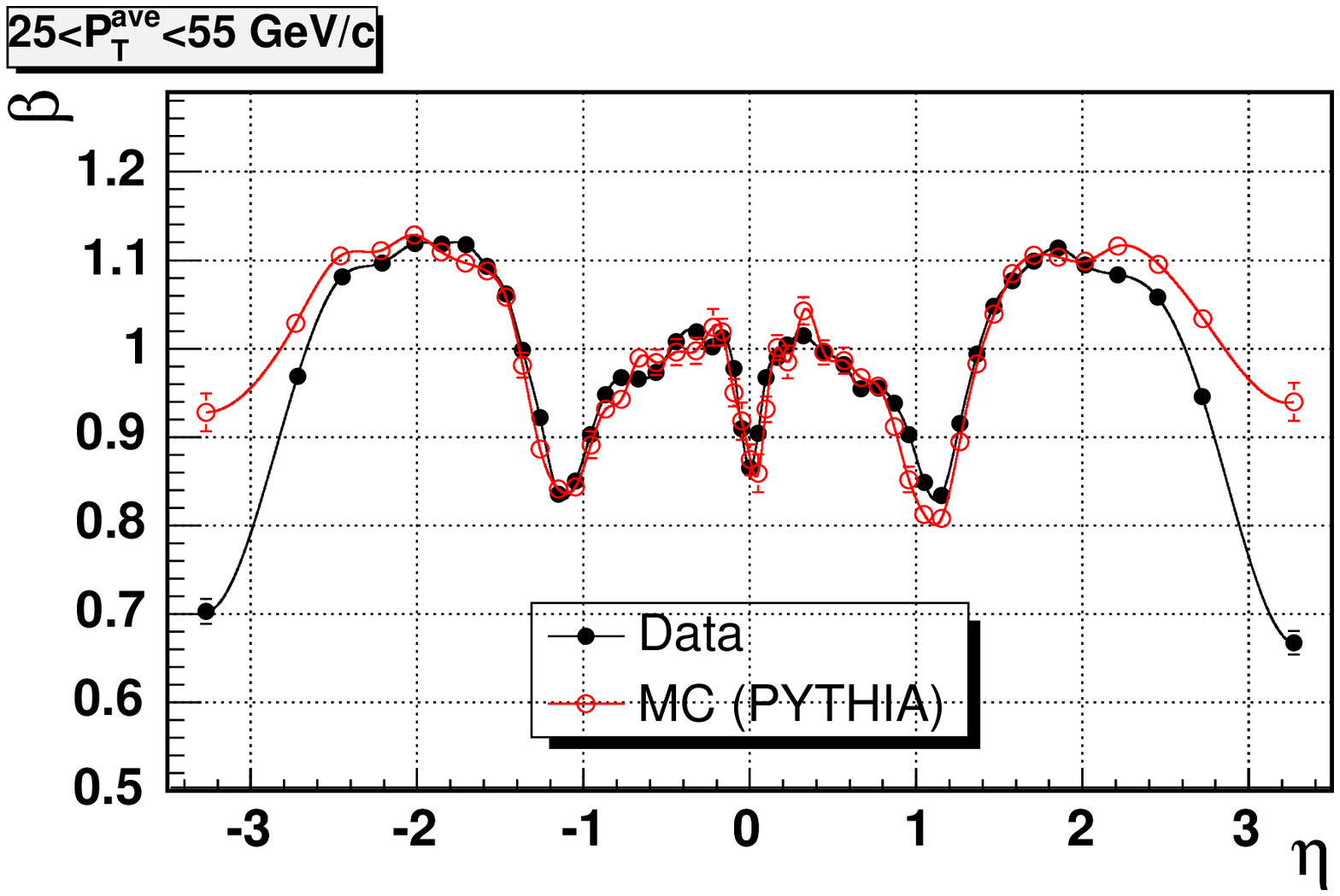}
  \caption{\label{fig:nim_status} 
  {\em Left:} Jet energy scale uncertainties due to
  calorimeter simulation and calibration published in~\cite{cdfnim}.
  {\em Right:} Di-jet balance $\beta$ vs. $\eta$ in
  data and simulation for transverse jet momenta
  25-55\,GeV/c. \vspace*{-9pt}}
\end{figure}

We first report on the performance of the early Run~II tuning embodied
by CDF physics publications to date. Despite the good agreement with
test beam data~\cite{currat:2002}, the quality of the simulation used
to be hard to control in the intermediate energy region, which is of
particular importance since it constitutes a relevant fraction of the
particle spectrum of a typical jet in CDF.  For the initial tuning of
the energy evolution of \gflash\ parameters, minimum bias data
providing tracks only up to 5\,GeV were involved. Later checks based
on special single track samples at higher momenta revealed a
underestimation of the data. The level of agreement in the central was
2\% for $<$12\,GeV, 3\% for 12 to 20\,GeV, and 4\% for
$>$20\,GeV~\cite{cdfnim}. Through convolution with a jet's typical
particle spectrum, these numbers directly translate into the systematic
uncertainties of the jet energy scale used by CDF to date
(Fig.\,\ref{fig:nim_status}, {\em left)}. The dominant contribution to
the total uncertainties originate from discrepancies between the
simulated and measured calorimeter response and from the low
statistical precision of early Run~II control samples.

The inhomogeneity of the calorimeter response is accounted for by an
$\eta$ dependent tuning. Jet responses in the plug and wall part are
re-calibrated w.r.t. the better understood central part using a
correction derived from di-jet events,
$\beta=p_\mathrm{T}^\mathrm{probe}/p_\mathrm{T}^\mathrm{trigger}$,
which relates the transverse momentum $p_\mathrm{T}$ of the
non-central ``probe'' jet to the $p_\mathrm{T}$ of the central
``trigger'' jet. A comparison of the simulated and measured di-jet
balance (Fig.\,\ref{fig:nim_status}, {\em right}) shows that the
tuning is reproducing many calorimeter particularities along $\eta$.

%%%%%%%%%%%%%%%%%%%%%%%%%%%%%%%%%%%%%%%%%%%%
% In Situ Tuning Progress
%%%%%%%%%%%%%%%%%%%%%%%%%%%%%%%%%%%%%%%%%%%%

\section{In Situ Tuning Progress} \label{sec:insitu}

The situation in the central calorimeter has now substantially
improved due to the development of dedicated single track triggers
with high momentum thresholds up to 15\,GeV/c. The samples collected
so far (a total of over 20\,M events) allow a precise monitoring of
the simulation performance in the energy range 0.5-40\,GeV. In
addition, steadily increased single track samples selected in minimum
bias events now allow a consistent tuning of the plug simulation up to
incident particle energies of 20\,GeV.

% Fig.: E/p in situ
\begin{figure}
  \begin{tabular}{cc} 
  \includegraphics[width=.46\textwidth]{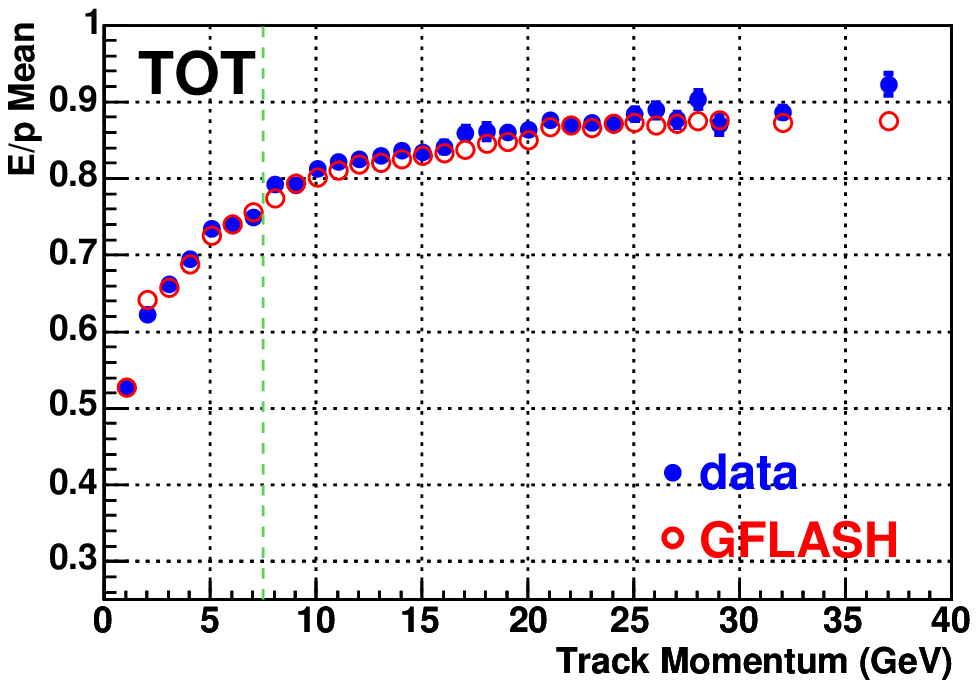} &
  \includegraphics[width=.46\textwidth]{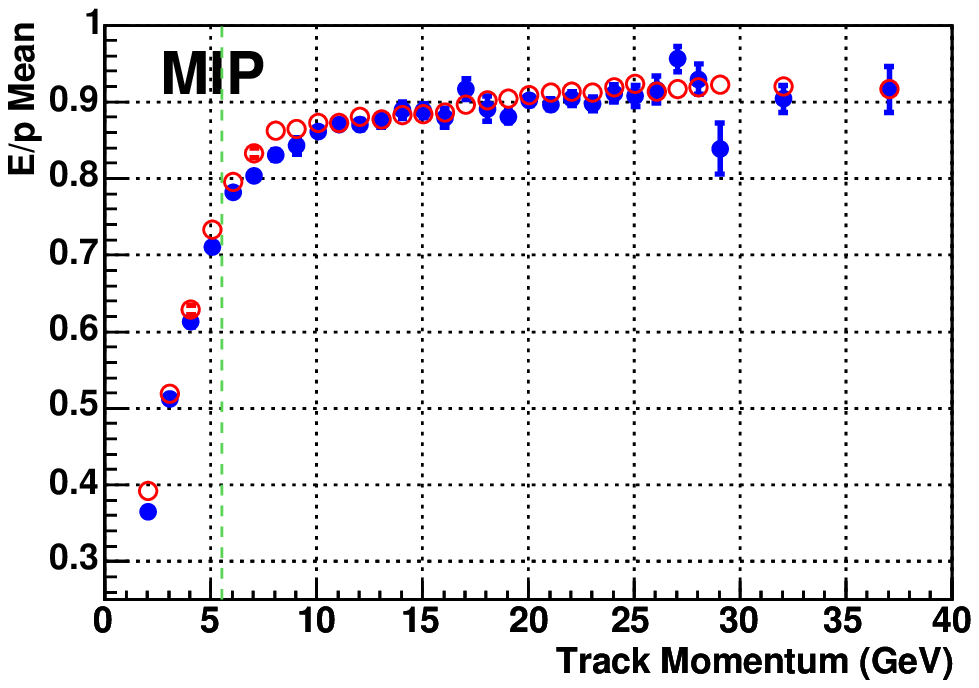}  \\[4pt]
  \includegraphics[width=.46\textwidth]{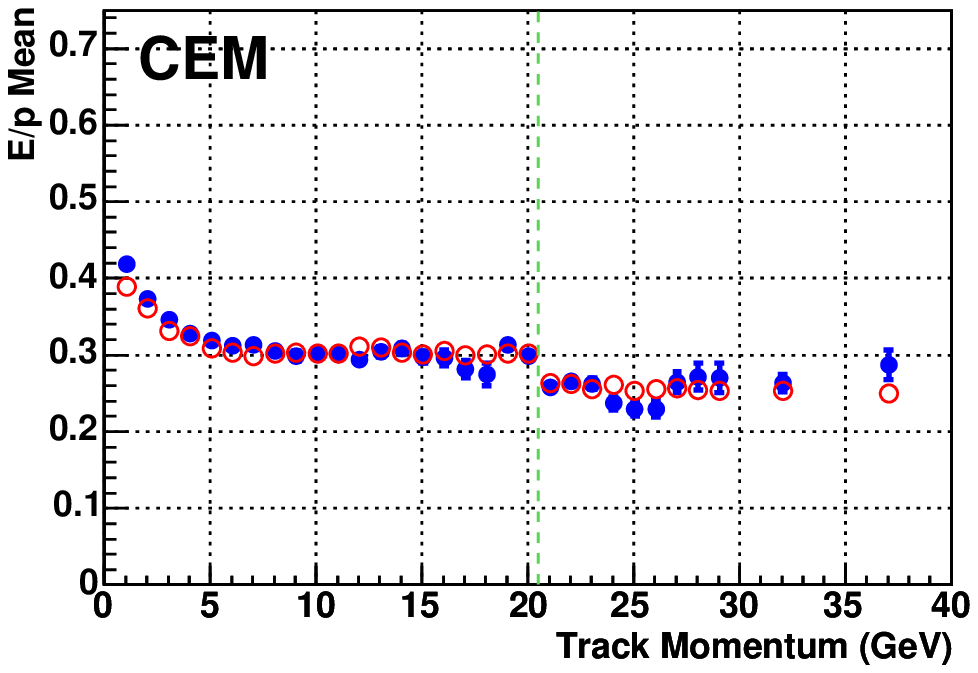} & 
  \includegraphics[width=.46\textwidth]{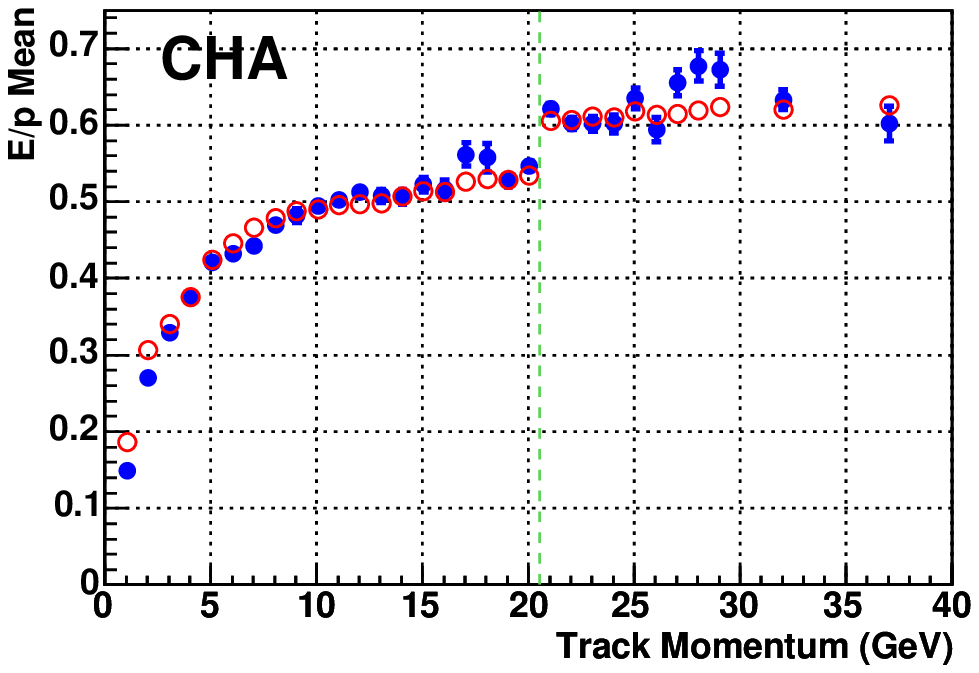}
  \end{tabular}
  \caption{\label{fig:insitu} Simulated and measured \EoP\ responses
  vs. track momentum in the central calorimeters CEM and CHA, the sum
  of both (TOT), and the CHA response of MIP-like particles (MIP). The
  vertical dashed lines indicate biases due to momentum dependent
  track analysis cuts and calculation methods. \vspace*{-8pt}}
\end{figure} 

\paragraph{Measurement technique}
The {\em in situ} approach employs high quality tracks which are well
contained within the target calorimeter tower (given by the track's
extrapolation into the electromagnetic compartment) and which are
isolated within a 7$\times$7 tower block around the target tower. The
signal is defined as the energy deposition seen in a 2$\times$2 block
(for PEM and CEM) or a 3$\times$3 block (for CHA and PHA),
respectively.  For the background estimate, tower strips with the same
$\eta$ range but along the edges of a 5$\times$5 tower group around
the target tower are used.  The simulation is usually based on a
particle gun using a controlled momentum spectrum of a mixture of
pions, kaons and protons, plus a minimum bias generator on top for
realistic background modeling.

\paragraph{Absolute \EoP\ response}
Fig.\,\ref{fig:insitu} shows the comparison of the average $E/p$
responses between data and simulation based on the most recent tuning
of the energy dependence of $f_\mathrm{dep}$, $\hat{e}/\hat{m}$ and
$\hat{h}/\hat{m}$. The basic idea is to adjust simultaneously the
simulated total response (TOT) and the CHA response of MIP-like
particles with CEM$<670$\,MeV using {\em in situ} tracks up to
40\,GeV, keeping the test beam tuning at higher energies. Indeed, the
simulated TOT and MIP {\em(top)} as well as the CEM and CHA responses
{\em (bottom)} agree quite well with the data. The total
\EoP\ in the central part, which serves as an important
benchmark for the systematic jet energy scale uncertainties, is
reproduced with 1-2\% precision within 0.5-40\,GeV. The new tuning
reflects more properly the current condition of the CDF calorimeter,
which includes aging effects of the photomultipliers, and partially
replaces the former conservative test beam uncertainties of 4\,\%.

% Fig: Lateral tuning
\begin{figure}
  \begin{tabular}{c}\\
  \includegraphics[height=.20\textheight]{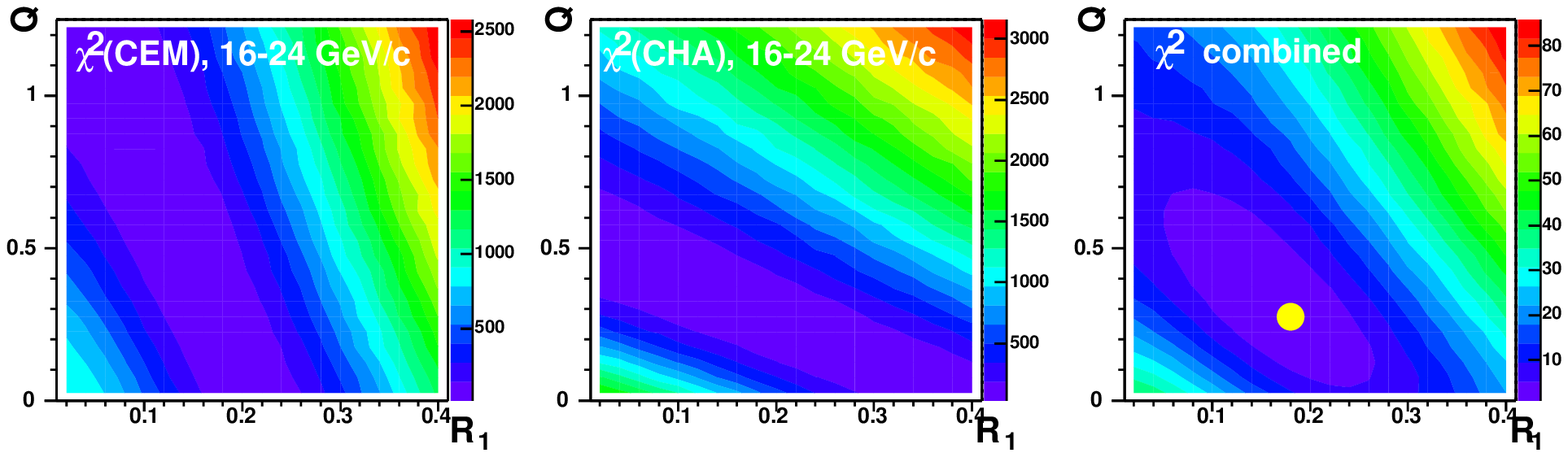} \\
  \includegraphics[height=.21\textheight]{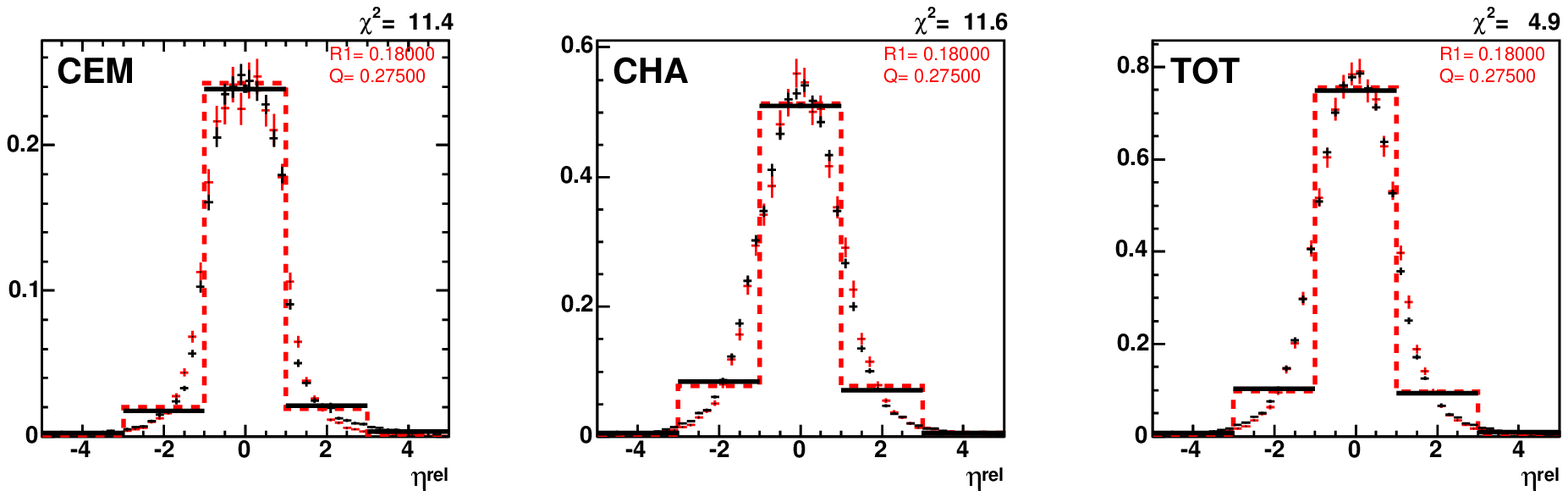} \\
  \end{tabular}
 \caption{ \label{fig:lat}{\em Top}:
 $\chi^2$ contours versus
 \gflash\ parameters $R_1$ and $Q$ (see text) obtained from
 comparisons of simulated and measured \EoP\ profiles in CEM (left)
 and CHA (center) for 16-24 GeV/c particles, and a weighted combination
 of both (right). {\em Bottom}: Simulated \EoP\ profiles versus data
 in CEM, CHA, and the sum of both (TOT) corresponding to the combined 
 minimum  $\chi^2$ position (highlighted point). \vspace*{-8pt}}
\end{figure}

\paragraph{\EoP\ profile} 
An accurate adjustment of the hadronic lateral profile is important
for two reasons. First, it controls the energy leakage out of the
limited \EoP\ signal regions. Thus, any profile mismatch between
simulation and data causes a bias when tuning the absolute
response. Second, leakage effects directly contribute to systematic
uncertainties due to the correction of jet energies for the energy
flow out of a jet cone~\cite{cdfnim}.

For the tuning, an experimental profile is defined using the
individual \EoP\ responses of five 1$\times$3 tower strips consecutive
in $\eta$, versus a relative $\eta$ coordinate normalized to the
$\eta$ of the target tower boundaries ($\eta=0$ denoting the center of
the target tower).  The availability of large single isolated track
samples allows a straightforward systematic approach: In
Eq.\,(\ref{eq:lat-b}), the constant $R_1$ denotes the shower core,
while $Q\equiv R_2 - R_3\ln E_\mathrm{inc}$ fixes how the profile
evolves with shower depth and incident particle energy. Since the
electromagnetic and hadronic calorimeter compartments probe different
stages of the average shower development, $R_1$ and $Q$ can be
constrained using the profiles measured in the {\em individual}
calorimeter compartments.  The top of Fig.\,\ref{fig:lat} shows a
comparison of 16-24\,GeV profiles between data and simulation in terms
of a standard $\chi^2$ estimator.  CHA and CEM provide different
contours of preferred parameter values, which helps to resolve the
ambiguity due to the strong anti-correlation of $R_1$ and $Q$ by using
a combination of both {\em(top right)}.  Generally a good agreement
between data and simulation is obtained {\em (bottom)}. Thus, $R_1$
has been fixed from 0.5 to 40 (20)\,GeV in the central (plug)
calorimeter, and $R_2$ and $R_3$ can be extracted from the linear
energy dependence of $Q$.

%%%%%%%%%%%%%%%%%%%%%%%%%%%%%%%%%%%%%%%%%%%%
%% Electron Response
%%%%%%%%%%%%%%%%%%%%%%%%%%%%%%%%%%%%%%%%%%%%

\section{Electromagnetic Response}

The tuning of electromagnetic showers in \gflash\ presents less
difficulty and is, although important, not detailed in this report.
The simulated electromagnetic scale, which has been set using electron
test beam data, was validated in Run~II at low momenta using electrons
from $J/\psi$ decays and at high momenta using electrons from $W/Z$
decays. The simulation reproduces the measured \EoP\ response with a
precision of 1.7\%. The uncertainty is dominated by a contribution of
1.6\% due to electrons pointing at the cracks between the towers,
whereas electrons well contained in the target tower account for less
than 1\%~\cite{cdfnim}. The crack response is complicated due to the
presence of instrumentation ({\em e.g.} wavelength shifter) but can be
monitored using electron pairs from $Z$ decays. One electron leg is
required to be well contained within a target tower and serves as an
energy reference, the other leg is used as probe to scan the $\EoP$
profile along the tower up to the crack. The description of the crack
response in \gflash\ has thus been improved using a correction
function applied to the lateral profile mapping of energy spots.

%%%%%%%%%%%%%%%%%%%%%%%%%%%%%%%%%%%%%%%%%%%%
%% Conclusions
%%%%%%%%%%%%%%%%%%%%%%%%%%%%%%%%%%%%%%%%%%%%

\section{Conclusion and Outlook}

\gflash\ has been tuned to reproduce the average hadronic response in the 
CDF central calorimeter with a precision of 1-2\% within the energy
range 0.5-40\,GeV/c.  The electron response is reproduced with similar
precision. An {\em in situ} tuning technique developed and re-fined in
Run~II has proved to be crucial to overcome past and current
performance limits. \gflash\ might also be a promising simulation tool
for LHC experiments. It is more flexible than \geant, it is tunable,
and it demonstrates excellent CPU performance.

%%%%%%%%%%%%%%%%%%%%%%%%%%%%%%%%%%%%%%%%%%%%%%%%
%% Acknowledgements
%%%%%%%%%%%%%%%%%%%%%%%%%%%%%%%%%%%%%%%%%%%%%%%%
\begin{theacknowledgments}

The main author would like to thank the German Max Planck Society and
the Alexander von Humboldt Foundation for their support, and is
grateful to the members of the CDF simulation group, in particular
S.Y.~Jun, S.A.~Kwang, Y.S.~Chung, G.~Yu and D.A.~Ambrose.

\end{theacknowledgments}

%%%%%%%%%%%%%%%%%%%%%%%%%%%%%%%%%%%%%%%%%%%%%%%%
%% Bibliography
%%%%%%%%%%%%%%%%%%%%%%%%%%%%%%%%%%%%%%%%%%%%%%%%

%\bibliographystyle{aipproc}   % if natbib is available
\bibliographystyle{aipprocl} % if natbib is missing

\end{document}